# Quantum-Inspired Genetic Algorithm for Designing Planar Multilayer Photonic Structure

Zhihao Xu[1], Wenjie Shang[1], Seongmin Kim[2], Alexandria Bobbitt[1], Eungkyu Lee[3, *], Tengfei Luo[1, 4, *]

1. Department of Aerospace and Mechanical Engineering, University of Notre Dame, Notre Dame, IN, 46556, USA

2. National Center for Computational Sciences, Oak Ridge National Laboratory, Oak Ridge, Tennessee 37830, USA.

3. Department of Electronic Engineering, Kyung Hee University, Yongin-si, Gyeonggi-do, 17104, Republic of Korea

4. Department of Chemical and Biomolecular Engineering, University of Notre Dame, Notre Dame, IN, 46556, USA

* Corresponding author. Email: Eungkyu Lee, eleest@khu.ac.kr; Tengfei Luo, tluo@nd.edu

**Abstract:** Quantum algorithms are emerging tools in the design of functional materials due to their powerful solution space search capability. How to balance the high price of quantum computing resources and the growing computing needs has become an urgent problem to be solved. We propose a novel optimization strategy based on an active learning scheme that combines the improved Quantum Genetic Algorithm (QGA) with machine learning surrogate model regression. Using Random Forests as the surrogate model circumvents the time-consuming physical modeling or experiments, thereby improving the optimization efficiency. QGA, a genetic algorithm embedded with quantum mechanics, combines the advantages of quantum computing and genetic algorithms, enabling faster and more robust convergence to the optimum. Using the design of planar multilayer photonic structures for transparent radiative cooling as a testbed, we show superiority of our algorithm over the classical genetic algorithm (CGA). Additionally, we show the precision advantage of the RF model as a flexible surrogate model, which relaxes the constraints on the type of surrogate model that can be used in other quantum computing optimization algorithms (e.g., quantum annealing needs Ising model as a surrogate).

## Introduction

The development of radiative coolers has emerged as a significant area of research in recent years since they can cool buildings and automobiles[1-3] by efficiently emitting thermal photons at the atmospheric window. Among them, transparent radiative coolers (TRCs) are ideal for window applications, where visible lights can pass through while non-visible solar light should be blocked to prevent excessive heating of the room or compartment. Recent innovations in this field have led to the exploration of new materials and the design of coatings that enhance the TRC's ability to reflect infrared solar radiation while remaining transparent to visible light[4-7]. Through radiating heat to the cold outer space via the atmospheric window, TRCs can achieve passive cooling without the need for external energy, thereby substantially reducing energy consumption[8-11]. In light of their planar geometry, precisely controllable dimensions, flexibility in combining different materials, and potential scalability in manufacturing, Planar Multilayer (PML) thin film structures are an ideal candidate platform for TRC development[8]. These PML structures have great potential in applications involving cooling buildings[8,12-14] and enhancing the efficiency of photovoltaic modules[15-20]. An optimized PML allows visible light to pass while reflecting other unnecessary spectral bands in solar radiation, thereby minimizing the solar heating effect to the greatest extent possible. For this, PML structures can consist of dielectric materials with distinctive refractive indices, and a spectral response across the solar wavelength range can be determined by the special combination and order of the material layers in the PML structure.

The design of PML can be formulated as an optimization problem, with the candidate material compositions as the input and the performance metric as the objective[21]. However, in many cases, numerical solutions to the resultant optimization problems become exceptionally challenging due to the myriad of local optima present in the objective function. In fact, the propensity of all optimization methods to converge at local minima means that searching for the global minimum or even decent local minima is a formidable task. Researchers have proposed a variety of optimization strategies for exploring potential optimal PML structures. Tikhonravov et al.[8,21,22] utilized needle optimization for the design of optical layers. However, needle optimization can be prone to falling into local optima[23]. Li et al.[24] and Shi et al.[25] employed the memetic method for optical designs. More recently, Wang et al.[26] used deep reinforcement learning to discover high-performance designs that outperform the structures designed by human experts and memetic algorithms. With the emergence and development of quantum computing, a variety of sophisticated, high-performance quantum algorithms have been employed to tackle optimization problems. Kitai et al.[27] and Kim et al.[28] utilized quantum annealing (QA) and factorization machine (FM) for the design of PML. These algorithms, based on quantum annealing, deliver a superior optimization speed and precision compared to classical computers. They provide highly accurate solutions for a specific kind of combinatorial optimization known as Quadratically Constrained Binary Optimization (QUBO)[29-31]. For PML design problems that have complex optimization landscapes, utilizing QUBO as the surrogate model and solving it with QA has been the necessary strategy since QA can only work with QUBO. While QA can serve as a powerful and reliable optimizer to find global optima of corresponding QUBO problems, the errors introduced by ignoring higher-order terms beyond the second order in the structure-performance relationship require many iterations between QUBO training/retraining, QA, and data collection before convergence is achieved[28].

One of the attempts to overcome the issues of premature convergence and tendency to be trapped in the local optima, prevalent in many optimization algorithms, is the Genetic Algorithm (or classical Genetic Algorithm, CGA), which effectively handles continuous solution exploration even with limited information[32]. However, this process is overly reliant on certain parameters, and the convergence rate can sometimes be slow. Motivated by these factors, the Quantum-inspired GA (QGA) was proposed.

In this work, we introduce an optimization algorithm based on a QGA for binary combinatorial optimization problems. As a test case, we use it to design PML for TRC applications. This optimization algorithm can utilize general surrogate models and find the optimum by QGA. We choose the Random Forest (RF) as the surrogate model in this study. Numerical experiments are conducted for structures with 6 to 20 layers, and the results show that our QGA-facilitated optimization algorithm can converge to comparable solutions as QA and overperforms classical genetic algorithm (CGA) on both convergence speed and global search capability. Furthermore, due to the advantages brought by the RF model, fewer iterations are needed for the QGA to converge to the optimal solution compared with QA-facilitated optimization. The proposed method provides a powerful tool for solving binary combinatorial optimization problems with complex searching spaces.

**Results**

**The Quantum Genetic Algorithm**

The QGA stands out with its smaller population size, rapid convergence speed, robust capability for global optimization, and strong resilience to variations in problem specifications. In QGA, a quantum chromosome, comprised of multiple qubits, represents a potential solution. It is noted that QGA mimics the evolution of physical qubits undergoing quantum logic gates, and the terminology of qubits and quantum gates in QGA, respectively, stands for the mathematical-quantum information and operators in the complex Hilbert space. The algorithm manipulates these chromosomes using quantum gates, drawing parallels to the genetic operations of crossover and mutation. Moreover, qubits can exist in a superposition of states, allowing exploration of a large search space and enhancing the global search capability of the algorithm. In addition, the evolution of quantum chromosomes can save memories by avoiding maintaining groups of binary vectors in CGA, which represent the population in the generation. QGA also utilizes the concept of quantum entanglement, enabling a more effective exchange of information between solutions. This makes QGA more adept at avoiding premature convergence to local optima, which is a common pitfall in CGA. In the evolution of QGA, the quality of the quantum chromosome is evaluated by calculating the fitness values of independent measurements of the quantum chromosome (the details of quantum measurement are in the Method section). The fitness values are evaluated by a figure-of-merit (FOM), which is used to describe how close between a designed PML structure and the ideal TRC (the details of the FOM definition are in the Supplementary Information, Section 1). the fitness value $f = -100 \times$ FOM. The measurements with high fitness values provide genetic information for the evolution direction of quantum chromosomes.

Compared with the CGA, QGA employs the probability amplitudes of qubits to encode chromosomes (as depicted in Eqn. 11) and utilizes quantum rotation gates to execute the chromosomal updated operations. Consequently, the formulation of the quantum rotation gates serves as a critical aspect of QGA, directly influencing the algorithm's performance. Here, we apply rotation-$Y$ ($Ry$) gates on each qubit in the chromosome to realize the evolution. The $Ry$ gate can be defined in a matrix form as[33-35]:

$$Ry(\theta) = \begin{bmatrix} \cos\frac{\theta}{2} & -\sin\frac{\theta}{2} \\ \sin\frac{\theta}{2} & \cos\frac{\theta}{2} \end{bmatrix} \quad (1)$$

The updated process is:

$$\begin{bmatrix} a_k^{t+1} \\ b_k^{t+1} \end{bmatrix} = Ry(\theta_k^t) \begin{bmatrix} a_k^t \\ b_k^t \end{bmatrix} = \begin{bmatrix} \cos\frac{\theta_k^t}{2} & -\sin\frac{\theta_k^t}{2} \\ \sin\frac{\theta_k^t}{2} & \cos\frac{\theta_k^t}{2} \end{bmatrix} \begin{bmatrix} a_k^t \\ b_k^t \end{bmatrix} \quad (2)$$

where $[a_k^{t+1}, b_k^{t+1}]^T$ and $[a_k^t, b_k^t]^T$ represent the probability amplitudes of the $k$-th qubit in the chromosome at the $t+1$-th generation and the $t$-th generation, respectively. $\theta_k$ is the rotating angle for the $Ry$ gate. We used an adaptive adjustment strategy[33] for the quantum rotation angle so that the direction and magnitude are updated dynamically during the evolutionary process. A large rotating angle is set early in the evolutionary process to quickly explore the solution space and find regions likely to contain optimal values. To accurately find the optimal value, we reduce the magnitude of rotating angle as evolution continues so that it eventually converges to a global optimum. The update of rotating angle follows the following formula:

$$\theta_i = \theta_{max} - \left(\frac{\theta_{max} - \theta_{min}}{N}\right) \times i \quad (3)$$

where $\theta_i$ is the value of the rotating angle of the $i$-th generation, $N$ is the max generation, $\theta_{max}$ and $\theta_{min}$ is the upper bound and lower bound of $\theta$. The adjustment strategy of rotating angle is shown in Table 1[36].

**Table 1:** Adjustment Strategy of Rotating Angle.

| $x_i$ | $best_i$ | $f(x) > f(best)$ | $\theta$ | $s(a_i, b_i)$ | | | |
|---|---|---|---|---|---|---|---|
| | | | | $a_i b_i > 0$ | $a_i b_i < 0$ | $a_i = 0$ | $b_i = 0$ |
| 0 | 0 | False | 0 | 0 | 0 | 0 | 0 |
| 0 | 0 | True | 0 | 0 | 0 | 0 | 0 |
| 0 | 1 | False | $\theta_i$ | +1 | −1 | 0 | ±1 |
| 0 | 1 | True | $\theta_i$ | −1 | +1 | ±1 | 0 |
| 1 | 0 | False | $\theta_i$ | −1 | +1 | ±1 | 0 |
| 1 | 0 | True | $\theta_i$ | +1 | −1 | 0 | ±1 |
| 1 | 1 | False | 0 | 0 | 0 | 0 | 0 |
| 1 | 1 | True | 0 | 0 | 0 | 0 | 0 |

\* $x_i$ and $best_i$ represent the $i$-th bit of chromosomes and best individual, respectively.
\*\* $f(\cdot)$ is the fitness evaluation function. The best individual has the highest fitness.
\*\*\* $s(a_i, b_i)$ is the direction of the rotating operation. +1 represents clockwise rotation.

In the CGA, the crossover operation in the evolution allows for an expansive exploration of the solution space. However, the quantum chromosome itself has the property of individual diversity resulting from quantum superposition. So, there is no need to perform the crossover operation in the QGA. On the other hand, mutation is the operation that can ensure sufficient variety in the population to avoid local optima, which is important for both CGA and QGA. Different from CGQ, quantum mutation will appear on the quantum chromosome and completely reverse the individual's evolutionary direction by swapping the value of probability amplitudes *a* and *b* of mutated qubits, which is implemented by *X*-Gate on the randomly selected qubits as[34]:

$$X = \begin{bmatrix} 0 & 1 \\ 1 & 0 \end{bmatrix}, \quad X \begin{bmatrix} a_k^t \\ b_k^t \end{bmatrix} = \begin{bmatrix} b_k^t \\ a_k^t \end{bmatrix} \tag{4}$$

Additionally, at the start of each QGA iteration, the genetic information of the optimal individual obtained from the previous QGA iteration will be partially incorporated into the initialization of the quantum chromosome, which reinforces the retention and utilization of valuable genetic information, leading to more efficient convergence towards optimal solutions. Therefore, the initialization of the quantum chromosome $q$ in iteration $\tau$ is:

$$\begin{aligned} q(\tau = 0) &= \left[ Ry(\theta_k^{rand}) \cdot \begin{bmatrix} \frac{1}{\sqrt{2}} \\ \frac{1}{\sqrt{2}} \end{bmatrix}_k \quad \text{for } k = 1,2,\dots,n \right] \\ q(\tau > 0) &= \left[ w_p \cdot \begin{bmatrix} a_k \\ b_k \end{bmatrix}^{it-1} + (1-w_p) \cdot Ry(\theta_k^{rand}) \cdot \begin{bmatrix} \frac{1}{\sqrt{2}} \\ \frac{1}{\sqrt{2}} \end{bmatrix}_k \quad \text{for } k = 1,2,\dots,n \right] \end{aligned} \tag{5}$$

where $\theta^{rand}$ is the randomly generated initial angle between 0 to π, $w_p$ is the weighing factor to balance the genetic information from prior QGA iteration and random initialization in the current iteration. Each qubit is firstly initialized to a uniform superposed state by the Hadamard gate as:

$$H = \frac{1}{\sqrt{2}} \begin{bmatrix} 1 & 1 \\ 1 & -1 \end{bmatrix}, \quad H \begin{bmatrix} 1 \\ 0 \end{bmatrix} = \begin{bmatrix} \frac{1}{\sqrt{2}} \\ \frac{1}{\sqrt{2}} \end{bmatrix} \tag{6}$$

Then, a series of *Ry* gates with randomly generated angles is applied to each qubit to produce a random quantum chromosome for QGA. Except for the first QGA iteration, other iterations will start with the initialized quantum chromosome which also incorporates the solution of former QGA iteration with weighing factor $w_p$ as shown in Eqn. 5. The determination of $w_p$ is based on the following formula:

$$w_p = \frac{1}{2} \frac{best_{current}}{best_{all}} \tag{7}$$

where $best_{all}$ and $best_{current}$ are the best fitness values of all time and in the current iteration, respectively. If the current iteration successfully converges to or exceed the fitness value of all time, $best_{all}$ will equal to $best_{current}$ at the end of the iteration, so that $w_p = 0.5$. Otherwise, $best_{all}$ will be higher than $best_{current}$ at the end of the iteration, which means that the initialization of the next iteration will have higher degree of randomness to encourage exploration of the entire landscape.

Additionally, to further mitigate the risk of the algorithm falling into a local optimum, we introduce an operation which we call memory corruption. This operation allows the optimal result, inherited from the previous cycle, to be discarded with a certain probability. This ensures that the incoming iteration does not simply inherit the optimal solution from the previous one. Consequently, an observable decay of the best solution can occur during the evolution of QGA, preventing stagnation and encouraging the search for new potential solutions.

The evolution of the QGA will terminate at a pre-defined maximum generation. However, if the evolution remains stagnant for a long time, it will be considered to have reached a local optimum and thus will be terminated. The termination criteria can vary based on the specific optimization problem. For the optimization problem discussed in this paper, we have determined that if the QGA evolution remains in a certain state for more than half of the maximum generation, the evolution will be terminated automatically.

In summary, the improved QGA process is described in Fig. 1.

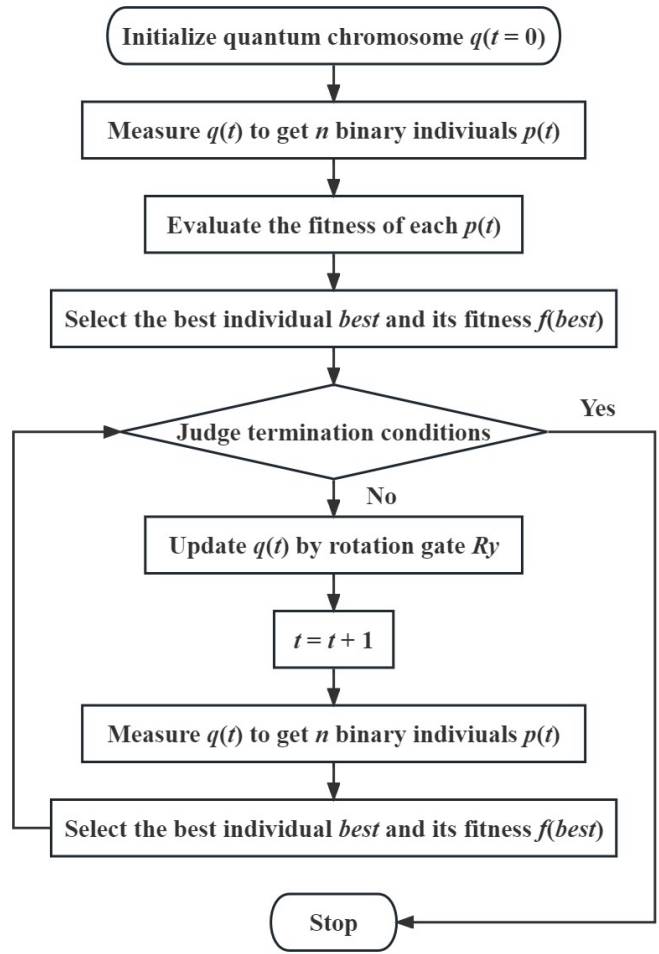

**Figure 1.** The Workflow of QGA

**Optimization of PML TRC**

We consider the design of an $N$ layer PML for TRC (as shown in Figure 2(a)). Each layer of the PML structure can be one of four candidate materials: silicon dioxide ($SiO_2$), silicon nitride ($Si_3N_4$), titanium dioxide ($TiO_2$), or aluminum oxide ($Al_2O_3$), and each material is assigned one of the combinations of two binary labels (00 = $SiO_2$, 01 = $Si_3N_4$, 10 = $TiO_2$, 11 = $Al_2O_3$). The concatenation of these binary labels in order gives a $2N$-long vector representing the structure. The optimization of PML structures is to minimize the FOM. The perfect PML can block all UV and IR light while allowing all visible light to transmit through. The proposed QGA, as an optimizer, is utilized to perform the optimization. As a comparison, we have also used CGA and QA for the same optimization problem. The details of QA can be found in Supplementary Information, Section 3.

Figure 2(b) shows the schematic of the QGA-facilitated optimization algorithm proposed for PML optimization. The algorithm is based on active learning scheme with iterations of random forest (RF) training, QGA optimization, and TMM calculations. The details of the algorithm can be found in Method section.

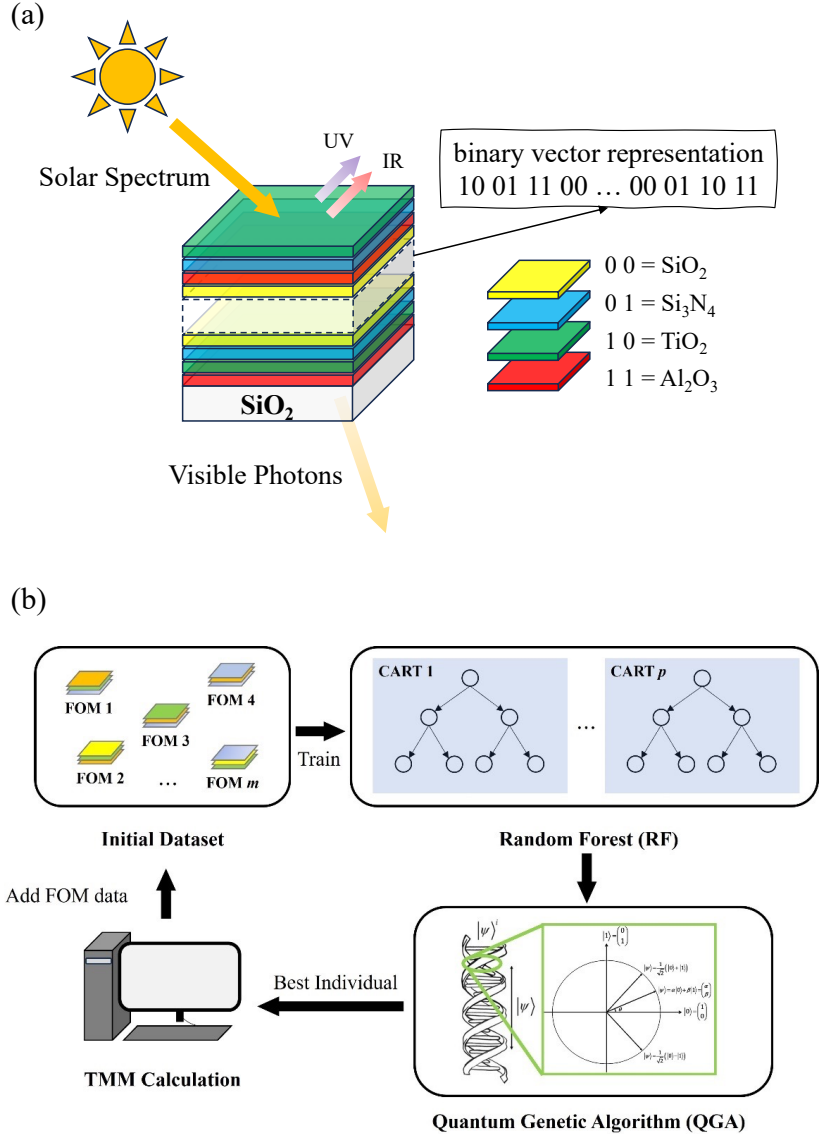

**Figure 2.** (a) The schematic structure of the PML in the TRC that can be mapped into a binary vector. (b) The workflow of active learning iteration between RF, QGA and TMM.

We first take $N = 6$ as a benchmark study to test our algorithm, for which we can afford performing an exhaustive search by calculating the optical properties using the transfer matrix method (TMM), which is the most efficient method to calculate optical characteristics of PML structures and evaluate the FOM of every one of all the $4^6 = 4,096$ possible structures. The optimal structure, denoted as [10 10 00 10 01 11], is identified with a FOM of 1.7713 from this brute force exhaustive search. We then used our QGA-facilitated optimization to solve the same $N = 6$ problem. The QGA-facilitated optimization is implemented with an initial training set with the amount of data of $m = 25$ and a maximum iteration number of 10. In each generation, the quantum chromosome undergoes 25 measurements, and fitness evaluations are performed on these measurements using the RF surrogate model to discern the evolution direction of the quantum chromosome. An adaptive rotating angle, as defined in Eqn. 3, decays from the upper to the lower bounds of $0.1\pi$ and $0.01\pi$, respectively. Throughout the evolution process, the mutation rate is maintained at 0.001. The QGA terminates after 100 generations or if it reaches a converged state before that. Figure 3 shows the evolution of FOM in the QGA optimization. The blue curve, orange curve and green curve represent the best FOM in the current generation, the average FOM in the current generation and the best FOM of all time, respectively. The $x$-axis is the step in the entire optimization, and each iteration of the active learning has 100 steps (correspond to 100 generations in QGA). As the figure shows, in each iteration, the FOM decays to a converged value, and the next iteration starts with a newly initialized quantum chromosome which partially include the information from the last iteration (the details of the quantum chromosomes initialization is shown by Eqn. 5). After only 5 iterations, the FOM has successfully converged to 1.7713, which is consistent with the solution obtained from the exhaustive search. Compared with the 4096 TMM calculations required by the brute force exhaustive search, QGA only requires 5 TMM calculations to label the solution of each iteration before getting the global optima. This benchmark test indicates the effectiveness and efficiency of our QGA-facilitated active learning optimization scheme.

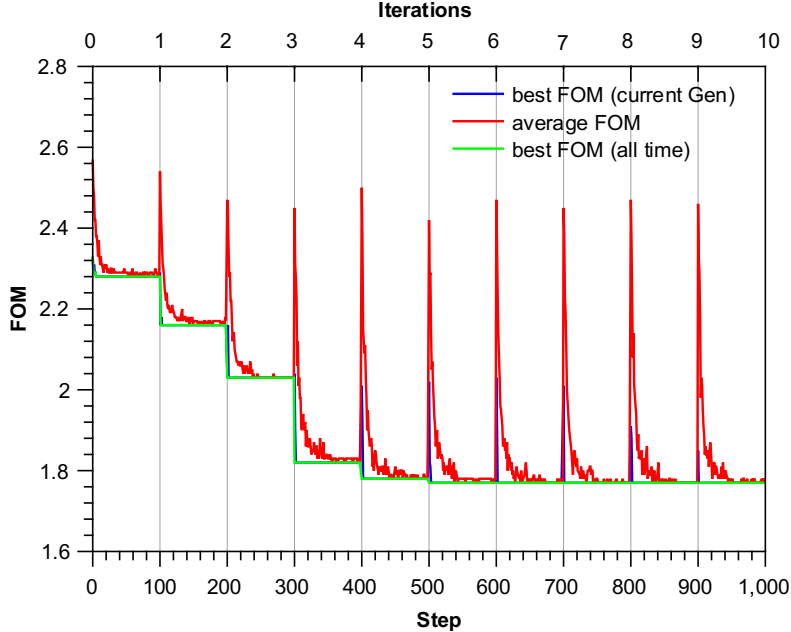

**Figure 3.** The evolution of FOM for $N = 6$ in QGA-facilitated active learning optimization.

The optimization is subsequently carried out for structures with $N = 8, 10, 16$ and 20. Optimizing structures with higher $N$ demands a larger population size and an increased number of generations in QGA. In addition, a larger initial dataset may also be needed to start the iterations efficiently. The parameters for the optimization are detailed in the Supplementary Information, section 2.

Figure 4 shows the evolution of the best FOM of all time for $N$ = 8, 10, 16 and 20. For low dimensional cases ($N$ = 8 and 10), our algorithm can discover accurate ground state obtained from exhaustive search faster than CGA-facilitated optimizations. For higher dimensional problems (such as $N$ = 16), our QGA algorithm can converge to the same structure as QA-facilitated optimization (using a real quantum annealer, DWave) within 7 iterations (under 5000 generations in total), while CGA start to fail obtaining a comparable result with QA. For the problem with $N$ = 20, both CGA and QGA algorithms cannot find the structure that is as good as QA finds. That is because the extremely large dimension of the search space requires an extremely large number of measurements in each generation to explore, which is prohibitively computationally expensive using a classical computer. Theoretically, the randomness introduced by quantum measurements, given a sufficient number of measurements, will guide the algorithm to explore the search space. The process involved in finding a local optimum, escaping from it, and then identifying a better solution, ultimately converging to a global optimum or a proper local optimum, can be very computationally intensive and time-consuming when using a classical computer to mimic quantum operations[37]. This would be easily affordable if real qubits are used and the state on the quantum computer can be easily prepared. However, there is currently no such a "large" quantum computer to access for a real application. Nevertheless, due to its inherent quantum state properties, QGA still outperforms CGA on both converging speed and exploring ability.

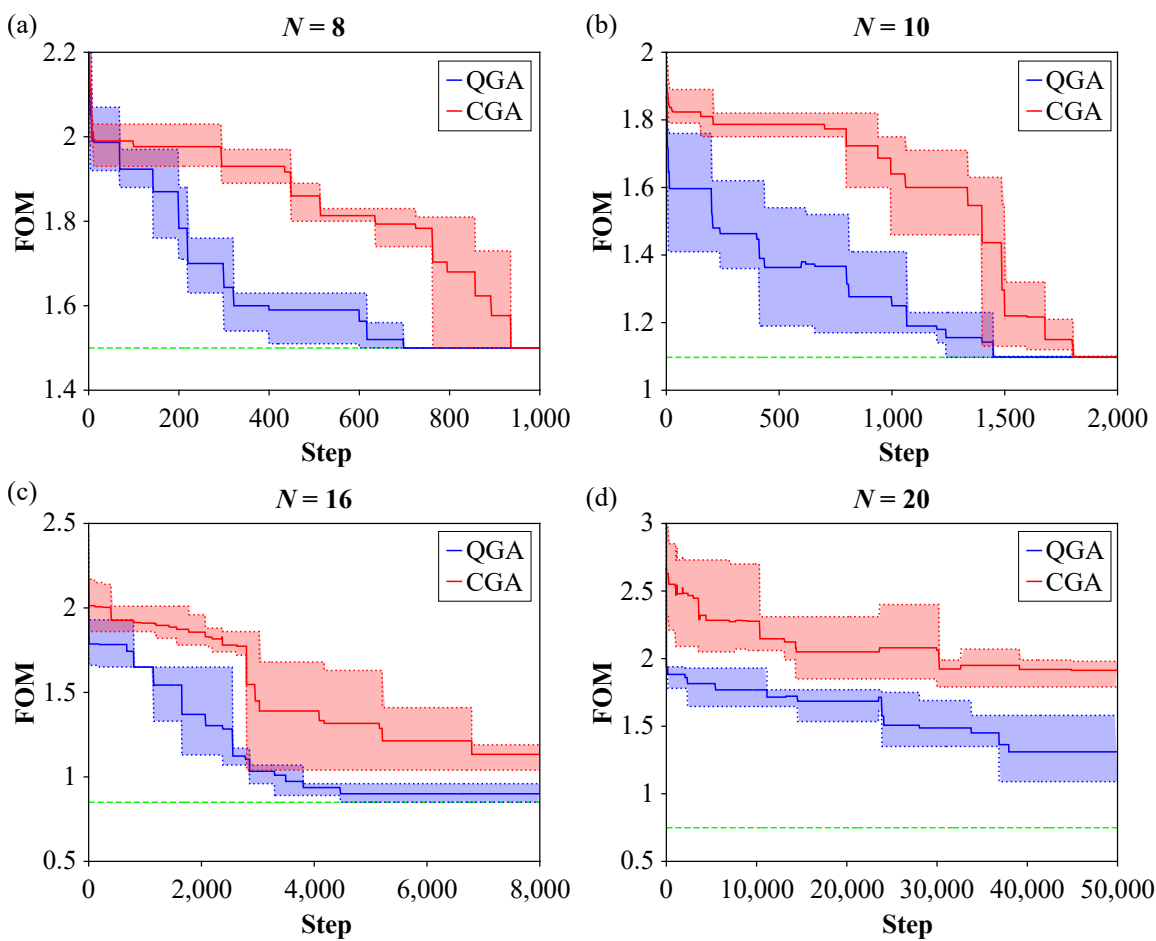

**Figure 4.** Evolution of the best FOM in QGA and CGA for (a) $N$ = 8, (b) $N$ = 10, (c) $N$ = 16 and (d) $N$ = 20. The solid lines represent the average values from 5 multiple independent numerical

experiments. The shadow regions represent the range of the best FOM from different trials. The horizontal green dash lines in the plots are the ground truth from exhaustive search (for $N = 8$ and 10) or the best solution from QA-assisted optimization algorithm (for $N = 16$ and 20).

**Computational Efficiency Analysis**

One crucial metric for evaluating the efficiency of an algorithm is its computational cost. In the context of the PML optimization using active learning, the total number of TMM calculations, which is the rate-limiting step in the iteration, serves as a key indicator of this cost. When optimizing an $N$-layered structure, an exhaustive search approach would necessitate $4^N$ TMM calculations to explore all possible solutions before identifying the true global optimum.

Figure 5 shows the relationship between the number of required fitness evaluations by TMM calculations and the number of layers in the design by using exhaustive search, QGA and QA. For QGA, only the cases that found the global optima are counted in comparison. Note, the $y$-axis is plotted on a log2 scale. The efficacy of our QGA-facilitated and QA-facilitated optimization is highlighted by their substantially reduced need for TMM calculations as compared to exhaustive search methods. Further amplifying this advantage is the high-performance nature of the RF model, which enables the QGA-facilitated optimization algorithm to require even fewer TMM calculations than its QA-based counterpart. This results in a significant alleviation of the burden traditionally associated with data acquisition, which is the TMM calculation in our case, but can be other physics-based modeling or experiments in other optimization tasks. This advantage is believed to be from the accuracy of the surrogate model, which is further discussed in the next section.

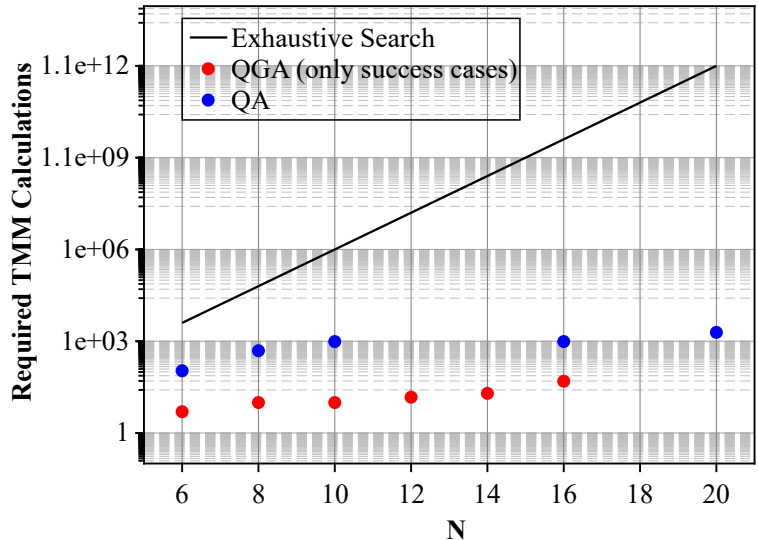

**Figure 5.** Comparison of computational cost for exhaustive search, QGA and QA.

**Comparison between random forest (RF) and factorization machine (FM) model**

Although the QGA's global search capability is currently not as good as QA in the high dimension optimization cases, the superiority of the RF model over factorization machine (FM), which is the required surrogate for QA as it can be mapped into a QUBO formulation, affords our proposed QGA-facilitated optimization algorithm a more reliable searching landscape for high-dimensional problems. Figure 6 provides a comparative analysis of the Rooted Mean Squared Error (RMSE) for FM and RF. Both are trained using TMM-calculated FOM data from of 8-layer, 10-layer, 16-layer, and 20-layer structures, with varying numbers of data points (25, 50, 75, and 100) randomly selected. 10 independent experiments have been performed and the standard errors of the mean are calculated as the error bars. Training is implemented with 5-fold cross-validation and Stochastic Gradient Descent (SGD), and RMSE is evaluated on a randomly selected test set. The results show that RF models outperform FM on the test sets. Moreover, as the data dimensionality increases, the advantages of RF become more apparent. From Figure 6, especially in panels (c) and (d), it can be seen that the RF model trained with less data can achieve higher accuracy than the FM model trained with more data, which indicates that the RF model requires less data to achieve higher prediction accuracy than FM model, which should be the reason why fewer iterations are necessary in our QGA-facilitated optimization algorithm than the QA-assisted optimization.

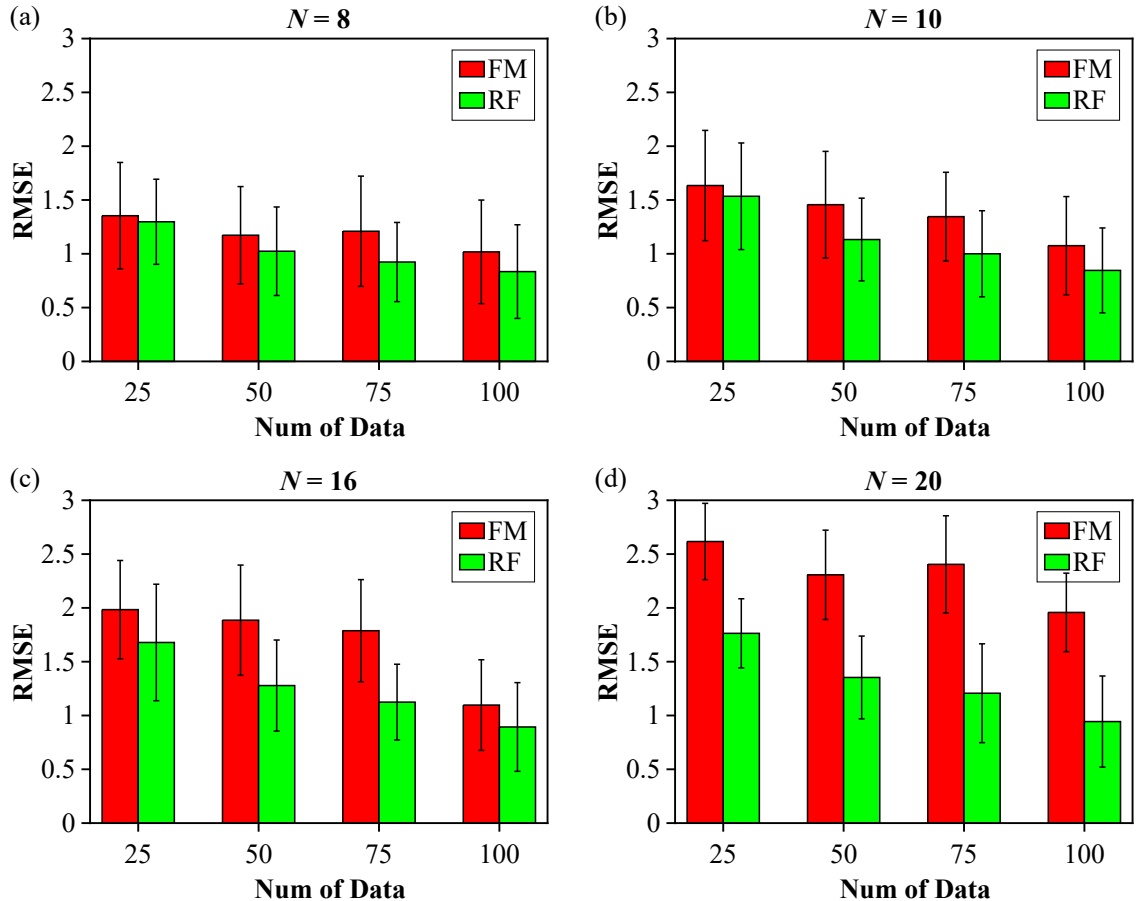

**Figure 6.** Comparison between FM and RF models prediction RMSE for $N = 8$, 10, 16 and 20. RMSE is evaluated on the same test dataset for FM and RF models. The error bars are calculated from the results of 10 independent experiments.

To further confirm this hypothesis, the trained RF and FM models are subsequently employed as surrogate models for the QGA in optimizing PML structures with dimensions of $N = 8$ and 16. Figure 7 illustrates the evolution curves of QGA when utilizing RF and FM as surrogate models with the same initialization. It can be observed that for the low-dimensional problem of $N = 8$, QGA always converges to the optimal structure regardless of the surrogate model employed, and the converging speed using RF is faster than FM. However, for the higher-dimensional scenario ($N = $ 16), employing RF as the surrogate model enhances the speed of QGA's convergence greatly. This substantiates the advantage of using RF as a more general surrogate model over FM.

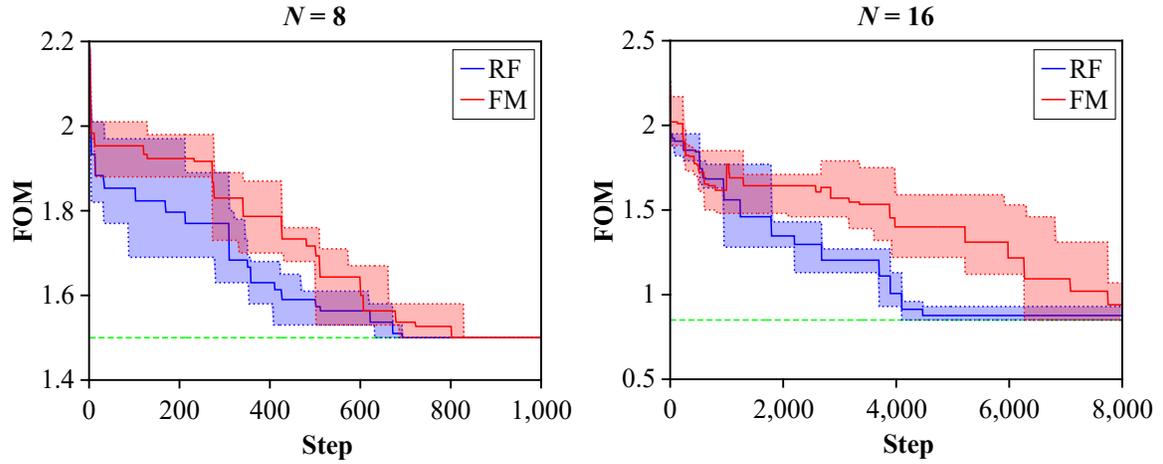

**Figure 7.** QGA evolution with RF surrogate model and FM surrogate model for $N = 8$ and $N = 16$. Solid lines and shadows represent average FOM evolution and range of multiple trials.

**Energy Saving Analysis**

Lastly, we estimated the benefit of the 16-layer and 20-layer TRC designed by our algorithm as a potential window material by calculating the energy it can save annually in the U.S. using EnergyPlus[28,38]. The annual energy saving over the surveyed U.S. cities is shown in Figure 8 (a) and (b). On average, the application of the 16-layer TRC as a window material would yield an annual energy saving of 33.58 MJ/m$^2$ over the surveyed locations. Even for the 20-layer TRC structure, despite not reaching the global minimum, can still contribute to an average annual energy saving of 26.03 MJ/m$^2$. Figure 8 (c) and (d) show the energy savings for the top 15 energy consuming states in the U.S. For the hot states (e.g., Arizona, Nevada, and Hawaii), our designed 16-layer TRC can potentially save ~30% of the cooling energy compared to conventional windows. The same calculations have also been performed for 20-layer TRC, and an energy saving over 20% can be achieved.

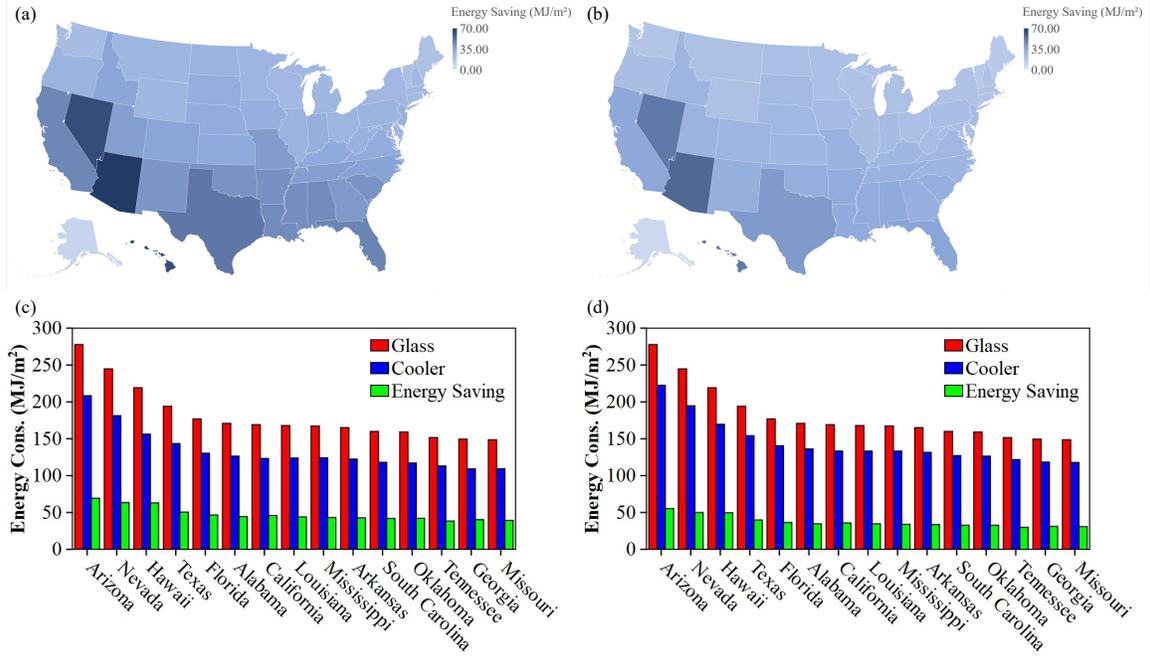

**Figure 8.** Estimated cooling energy saving across the U.S. by using the (a) 16-layer and (b) 20-layer TRC as the window material. The annual energy savings of the top 15 energy-consuming states over U.S. by applying (c) the 16-layer TRC and (d) the 20-layer TRC designed by our QGA-facilitated optimization.

**Discussion**

In this work, we introduce an optimization algorithm based on QGA to address the challenge of finding the optimal in binary combinatorial problems. We used PML for TRC as an application example. The combination of QGA and RF regression model iteratively operates in an active learning framework to obtain global optimum in complex and discrete search spaces. QGA excels in aspects like smaller population size, faster convergence speed, superior global search ability, and robustness, greatly outperforming conventional genetic algorithms. Moreover, compared to employing FM (QUBO) to describe the search space, the RF regression model demonstrates stronger reliability in high-dimensional problems. Hence, unless the dimensionality of the problem is unreachable for classical computers, QGA can be more efficient compared QA not to mention CGA. At the same time, the results in this work point out the bottleneck of current QA-facilitated optimization scheme and strongly indicate the necessity for developing FM that with 3$^{rd}$ or higher orders and accessible by quantum annealers.

## Methods

### Quantum Computing

Quantum computing, an intriguing application of quantum mechanics in algorithmic computation, significantly diverges from classical computing primarily due to its intrinsic parallelism. Unlike classical computing where systems exist in definite states, quantum computing operates on the principles of superposition and entanglement, allowing the system to exist in multiple states simultaneously[39-41]. The state of a quantum system is described by a probabilistic wave function, the square of which provides the probabilities for the possible states[33]. This attribute expedites computation in quantum systems by order of magnitude compared to classical systems.

The evolution of qubits through so-called quantum channels is carried out by unitary quantum gates that manipulate the qubits just like classical bits are manipulated by logic gates in a computer[42,43]. These unique characteristics of quantum computing – superposition, entanglement, and quantum gate operations – confer it with unparalleled computational power[44]. Integrating quantum principles into optimization algorithms offers potential improvements, enhancing traditional methods and aiding in complex problem-solving.

The basic unit for information storage is qubit in the quantum computer. Unlike classical bits that can be in a state of either 0 or 1, a qubit can exist in a state of superposition, representing 0 and 1 simultaneously. The super position state of a qubit can be expressed as follows:

$$|\varphi\rangle = a|0\rangle + b|1\rangle \qquad (8)$$

where, $|\varphi\rangle$ is the superposition state. $a$ and $b$ are complex numbers which denote the probability amplitudes of the corresponding ground state $|0\rangle$ or $|1\rangle$, and $|a|^2 + |b|^2 = 1$. Therefore, an $n$-qubits quantum register can store the quantum state which is the coherent superposition of $2^n$ ground states:

$$|\phi\rangle = \sum_{i=1}^{2^n} c_i |\phi_i\rangle \qquad (9)$$

where $c_i$ are the probability amplitudes that satisfying:

$$\sum_{i=1}^{2^n} |c_i|^2 = 1 \qquad (10)$$

If we use a binary vector to encode qubits on the polymorphic problem, the system with $n$ qubits can be expressed as follows[45]:

$$q = \begin{bmatrix} a_1 & a_2 & a_3 & \dots & a_n \\ b_1 & b_2 & b_3 & & b_n \end{bmatrix} \qquad (11)$$

where $|a_k|^2 + |b_k|^2 = 1$ ($k = 1, 2, \dots, n$), and each pair of probability amplitudes $[a_k, b_k]^T$ represent a qubit in the system. When the quantum system is measured, coherence will disappear, and the quantum system will collapse to a definite state $|\Phi_i\rangle$ with probability given by the squared magnitude of the respective probability amplitude $|c_i|^2$ according to Eqn. 9. The probability of qubit $k$ being measured in state $|0\rangle$ (or $|1\rangle$) will be $|a_k|^2$ (or $|b_k|^2$) as Eqn. 11 described.

**Random Forest Algorithm**

Decision trees, particularly the Classification and Regression Tree (CART) method, are known for their simplicity and interpretability in supervised learning[46,47]. To counter the potential overfitting problem of single decision tree, the Random Forest (RF) algorithm, which was proposed by Breiman et al.[48] in 2001, is always employed as an ensemble method to enhance model robustness and predictive accuracy. The RF algorithm operates by constructing a multitude of decision trees and averaging related predictions from individual trees, and therefore it will yield a more accurate final prediction than any individual tree could offer[49-51].

**Active Learning Scheme with RF and QGA**

We employ an optimization algorithm based on the active learning framework. To circumvent the computationally intensive TMM calculation, we implement a RF as a surrogate model, significantly accelerating the calculation speed of the figure-of-merit (FOM) per instance. Initially, the RF model is trained on $m$ data points calculated by the TMM. Due to the limited quantity of training data, the accuracy of the Random Forest is not expected to be high initially or whole optimization space. Therefore, an iterative process between QGA and RF training is implemented within the active learning loop. During each iteration, the FOM of individuals in the QGA is evaluated using RF. The FOM calculated by RF surrogate model is mapped to the fitness value in the QGA, and the optimal structure found by the QGA is put into the TMM calculations to calculate the true FOM, which added to the database to train a new RF model, thereby promoting the progression of iterations.

In the algorithm, the structures of the TRC are represented by a series of binary vectors, which are also the input of TMM and RF calculations. After training the RF model, the QGA starts with the quantum chromosome initialized according to Eqn. 5, and the fitness evaluation in the QGA evolution is implemented by the surrogated RF model. After obtaining the best solution from QGA, TMM is called to verify the FOM of the best solution and add the data to the training set to retrain the RF model. If the solution of QGA has been included in the training set, a randomly selected structure and its FOM is added to the training set to further diversify the applicability of the RF model in the optimization space. With the iterations ongoing, the accuracy of RF will be improved in the design space, by which the fitness evaluation will be more reliable for finding the best solution.

**Acknowledgement**

This research was supported by the Quantum Computing Based on Quantum Advantage Challenge Research (RS-2023-00255442) through the National Research Foundation of Korea (NRF) funded by the Korean government (Ministry of Science and ICT(MSIT)). This research also used resources of the Oak Ridge Leadership Computing Facility at the Oak Ridge National Laboratory, which is supported by the Office of Science of the U.S. Department of Energy under Contract No. DE-AC05-00OR22725. The authors also would like to thank the Notre Dame Center for Research Computing for supporting all the simulations in this work.

Notice: This manuscript has in part been authored by UT-Battelle, LLC under Contract No. DE-AC05-00OR22725 with the U.S. Department of Energy. The United States Government retains and the publisher, by accepting the article for publication, acknowledges that the U.S. Government

**Contribution**

Z.X., E.L. and T.L. conceived the idea; Z.X. designed and performed the QGA code for the research; W.S. and S.K. performed the QA for the research; A.B. performed the energy saving calculations; Z.X. wrote most parts of the manuscript; All authors discussed the results and commented on the manuscript.

**Data Availability**

Requests for data and materials should be sent to the corresponding authors or Z.X. (zxu8@nd.edu).

**Code Availability**

The underlying codes for this study are available from the corresponding author or Z.X. (zxu8@nd.edu) upon reasonable request.

51

Supplementary Information for

# Quantum-Inspired Genetic Algorithm for Designing Planar Multilayer Photonic Structure

Zhihao Xu, Wenjie Shang, Seongmin Kim, Alexandria Bobbitt, Eungkyu Lee*, Tengfei Luo*

* Corresponding author: Eungkyu Lee, eleest@khu.ac.kr; Tengfei Luo, tluo@nd.edu

**Supplementary Note 1. Figure-of-Merit (FOM) of TRC and Transfer Matrix Method (TMM)**

To optimize the performance of TRC, we use a figure-of-merit (FOM) as the objective function. The FOM describes how close the designed TRC is to the ideally perfect TRC in terms of wavelength-dependent optical property. The definition of FOM is:

$$FOM = \frac{10 \int_{\lambda_1}^{\lambda_2} \bigl(Tl(\lambda) - Tl_{ideal}(\lambda)\bigr)^2 d\lambda}{\int_{\lambda_1}^{\lambda_2} S(\lambda)^2 d\lambda} \tag{S1}$$

where $Tl(\lambda) = S(\lambda)T(\lambda)$ is the transmitted irradiance of the TRC, in which $S(\lambda)$ is the solar spectrum (air mass 1.5 global) and $T(\lambda)$ is the transmission efficiency of the TRC. $Tl_{ideal}(\lambda)$ is the transmitted irradiance of the ideal TRC, which has 0 transmission efficiency in the UV (< 400 nm) and IR (750 to 2500 nm) range and 1 in the visible range (400 to 750 nm). Minimizing the FOM will allow the designed TRC to approach the ideal TRC. The calculation of FOM is done by the transfer matrix method (TMM)[1].

The TMM is a method that has been widely used in optics and acoustics to analyze the propagation of electromagnetic or acoustic waves through a stratified medium. It is based on the fact that there are simple continuity conditions for the electric field across boundaries from one medium to another, which is indicated by Maxwell's equations. If the field is known at the beginning of a layer, the field at the end of the layer can be derived from a simple matrix operation. For the stack of layers which we are trying to design in this work, a system matrix, which is the product of the individual layer matrices, can be used to represent the entire system. Therefore, through TMM calculation, we can establish the mapping relationship between the layered structure represented by a binary vector and its corresponding FOM and find the best structure by minimizing the FOM.

**Supplementary Note 2. Parameters for Optimizations**

The parameters of the optimizations for different structures are shown in Supplementary Table 1. The total number of fitness evaluations by the RF model are calculated by multiplying the number of generations in one iteration, number of iterations and population size of each generation, which are also listed in the table for each case.

**Supplementary Table 1.** Parameters of Optimizations for different *N*.

| # of layers | # of initial dataset | # of generations | # of iterations | population size (measurements) | # of total fitness evaluation by RF |
|---|---|---|---|---|---|
| 6 | 25 | 100 | 10 | 25 | 25,000 |
| 8  | 25  | 100  | 10  | 50  | 50,000 |
| 10 | 50  | 200  | 10  | 50  | 100,000 |
| 12 | 100 | 200  | 15  | 50  | 150,000 |
| 14 | 100 | 400  | 20  | 100 | 800,000 |
| 16 | 150 | 800  | 50  | 100 | 4,000,000 |
| 20 | 150 | 1000 | 100 | 500 | 50,000,000 |

Supplementary Figure 1 shows the relationship between the number of total numbers of fitness evaluations by the RF model and the number of layers $N$.

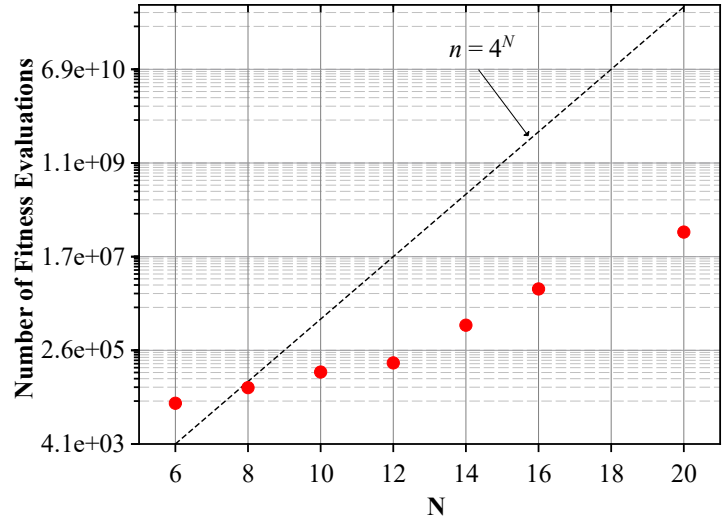

**Supplementary Figure 1.** Relationship between number of layers $N$ and number of total fitness evaluations by RF model.

**Supplementary Note 3. Quantum Annealing Assisted Optimization Algorithm**

To identify the optimal configuration for the multilayer coating, we adopted the methodology outlined by Kim et al.[2], harnessing quantum annealing-assisted iterative techniques. An initial dataset, comprising binary vectors representing the PML structure and their associated FOM, is curated and input into the Factorization Machine (FM) surrogate model. We operate under the assumption that the sampled PML structure accurately reflects the search space, and that the FM model can effectively learn this distribution from the training data. Consequently, optimizing the trained surrogate model should facilitate the discovery of the optimal PML structure with the lowest FOM. Subsequently, this high-dimensional FM is transformed into an appropriate QUBO format, which is then resolved using Quantum Annealing. The newly optimized PML configuration, as determined by quantum annealing, undergoes numerical validation via the TMM and is incorporated into the dataset for successive discovery cycles. The entire optimization procedure terminates either upon resource exhaustion or once convergence to a definitive structure is achieved.

All the QA procedures are run on the D-Wave quantum computer (Advantage System 4.1). Powered by the robust capabilities of quantum annealer and the superior of quantum annealing algorithm, the QA-assisted optimization, despite undergoing more iterations, is able to converge to the global optimum or good local optimum in a remarkably short physical time, and that is the reason why QA has been widely applied.

**Supplementary Reference**